\documentclass[showpacs,twocolumn,aps
]{revtex4-1}
\usepackage{graphicx}
\usepackage{url}
\usepackage[utf8x]{inputenc}

\newcommand{\beq}{\begin{equation}}  
\newcommand{\eeq}{\end{equation}}  
\newcommand{\bea}{\begin{eqnarray}}  
\newcommand{\eea}{\end{eqnarray}}  

\newcommand{\re}{Re}
\newcommand{\ecf}{E_{cf}}
\newcommand{\po}{\overline}

\usepackage{layouts}
\begin{document}

\title{Periodic orbits near onset of chaos in plane Couette flow}
\author{Tobias Kreilos} 
\affiliation{Fachbereich Physik, Philipps-Universit\"at Marburg, 
D-35032 Marburg, Germany}
\author{Bruno Eckhardt}
\affiliation{Fachbereich Physik, Philipps-Universit\"at Marburg, 
D-35032 Marburg, Germany\\
J.M. Burgerscentrum, Delft University of Technology, Mekelweg 2, 2628 CD Delft, The Netherlands}
\date{\today}

\begin{abstract}
We track the secondary bifurcations of coherent states in plane Couette
flow and show that they undergo a periodic doubling cascade
that ends with a crisis bifurcation. We introduce a symbolic dynamics
for the orbits and show that the ones that exist fall into the 
universal sequence described by Metropolis, Stein and Stein for unimodal
maps. The periodic orbits cover much of the turbulent dynamics in that their
temporal evolution overlaps with turbulent motions when projected onto a 
plane spanned by energy production and dissipation.
\end{abstract}


\maketitle\

{\bf The identification of the different 'routes to chaos'
\cite{Eckmann1981,Ott1981} has suggested ways in which they 
can also be 'routes to turbulence' that lead from
a laminar flow state through various bifurcations to a turbulent attractor. 
Studies of the turbulence transition in plane Couette flow and pipe flow, 
which do not have a linear instability of the laminar profile, have 
revealed a route to turbulence that passes through the formation of a chaotic
saddle \cite{Eckhardt2007,Eckhardt2008b}. The saddle develops around a scaffold of coherent structures that
appear in saddle-node type bifurcations, with the notable difference to the standard scenario that also 
the 'node'-states typically become unstable very quickly. In this transitional region the
turbulence is not persistent but transient, and can hence be associated
with a transient chaotic saddle rather than an attractor. In the spirit
of a dynamical systems approach to characterizing these features \cite{Gutzwiller1971,Gutzwiller1990,Cvitanovic1991}, we here
present an analysis of the state space structures near transition in 
plane Couette flow and study the various bifurcations and periodic orbits
that appear.
}

\section{\label{sec:intro}Introduction}

The application of dynamical systems ideas to fluid mechanical problems has provided much
insight into the ways fluids transit from a laminar state to a turbulent one. The various bifurcations that appear in the case of
fluids heated from below (Rayleigh-Benard) or fluids driven by centrifugal instabilities 
(Taylor-Couette) have been studied experimentally and theoretically in considerable detail \cite{Andereck1986,Koschmieder1993}.
Studies of turbulence in liquid helium have shown the period doubling cascade \cite{Maurer1979} and the
quasi-periodic routes to chaos
\cite{Stavans1985}, 
and the rich structures of secondary bifurcations and flow 
states in RB and TC have been explored \cite{Busse1978,Andereck1986,Koschmieder1993}. 
The approach has been so successful that flows without
linear stability of the linear profile, like plane Couette flow and pipe flow, have been
suspected to belong to the class of subcritical bifurcations
\cite{Cross1993}, 
perhaps with the bifurcation point at infinity (as reflected, e.g., in the title of Nagata's first papers
on coherent structures in plane Couette flow \cite{Nagata1990}). However, the first structures that are observed
near the transition point are not steady or simple flows, but they are fluctuating 
spatially and temporally, indicating that these states undergo many further bifurcations
as the Reynolds number is lowered.

A more promising approach starts with the turning points of the subcritical bifurcations,
the saddle-node bifurcations in which these structures appear. The possibility to approach
flows from this point of view is suggested by the remarkable persistence of key features
of these structures \cite{Eckhardt2007}: 
they are all dominated by downstream vortices which modify the
base profile to generate streaks. This by itself is not enough, as flows that are
translationally invariant in the downstream direction cannot be sustained \cite{Moffatt1990}.
So three-dimensionality is another key element, and also the reason for the 
absence of analytical studies of this transition scenario. 

The importance of vortices and streaks has been deduced from numerical studies of plane
Couette flow in small domains, where a recurrent pattern of vortices and streaks could be
identified \cite{Jimenez1991,Hamilton1995}.
The  interactions between vortices and streaks and the instabilities that streaks can undergo
has been studied by Waleffe \cite{Waleffe1995}, who proposed a self-sustaining process
(analyzed also in \cite{Schoppa2002, Lagha2007, Aubry1988, Holmes2005}).
This process is usually explained by starting with the vortices, which then
drive streaks, and the streaks then undergo an instability in which normal vorticity is 
created. 
The cycle is closed by nonlinear interactions that lead to the recreation of streamwise vortices.

An example where the transitions between the different states and their symmetry
can he analyzed directly, and in a physically meaningful
setting, is Taylor-Couette flow in the limit in which it approaches plane 
Couette flow \cite{Faisst2000}.
In the curved case, say with the inner cylinder rotating and the outer one at rest, the 
flow undergoes a first bifurcation to Taylor-vortex flow, in which vortices with an
azimuthal symmetry appear. As the rotation rate is increased, the vortices undergo a 
secondary bifurcation, in which an azimuthal modulation is added and the rotational
symmetry is broken. Increasing the radii while keeping the distance between the
cylinders fixed, the influence of the curvature becomes smaller and the flow approaches plane 
Couette flow. Along the way, the bifurcation structure
changes: the transition point to the formation of azimuthally symmetric vortices moves out
infinity, so that the linear stability of plane Couette flow is recovered \cite{Faisst2000}. The three-dimensional
wave structures, arising in secondary bifurcations of the Taylor vortex flow,
can be traced over to plane Couette flow, where they now appear
in a saddle node bifurcation \cite{Faisst2000}. In going from the Taylor-Couette to 
plane-Couette flow the transition can be studied within realistic models. In the case of 
pipe flow no physical family of flows could be identified, 
so a body force was introduced to embed the system into one where the bifurcations
can be realized. Despite this artificial setting, the coherent structures thus 
found are similar to the ones in  plane Couette flow and share many properties 
\cite{Faisst2003}.

The coherent structures just described are usually unstable, and can appear 
only transiently in the flow \cite{Hof2004,Schneider2007a,Lai2011}. 
Moreover, it was found that they are not sufficiently entangled to close off 
the state space around them to form an attractor: thus it is not only the 
coherent structures that are visited transiently, but the entire turbulent
dynamics becomes a transient phenomenon, characterized by an 
exponential distribution of lifetimes, as can be expected for a chaotic saddle
\cite{Hof2006,Schneider2008a}.
While this sets the stage from the dynamical systems point of view, several
questions remain to be addressed \cite{Eckhardt2008b}. We will address two of
them, related to the secondary bifurcations of coherent structures, in the following.

It was noticed long ago in \cite{Clever1997}, that
the coherent structures that appear in plane Couette flow at Re$=125$ are stable very 
close to the bifurcation point and become unstable via Hopf bifurcations. This would 
suggest that the transition creates an attractor rather than a transient saddle. However,
at higher Re the flowed evidently is transient: so how does the state space near these
coherent states change as one increases the Reynolds number and how does the flow
change from chaotic but persistent to transient? The analysis of these properties
in pipe flow shows that the sequence of bifurcations that lead from a
(small) chaotic attractor to transient dynamics can be rather intricate
\cite{Mellibovsky2011,Mellibovsky2012}.

A second item concerns the possibility to describe the flows using periodic orbits,
as is possible in other dynamical systems \cite{Cvitanovic1988,Cvitanovic1991,Eckhardt1994,Christiansen1997}. 
Identifying periodic orbits in high-dimensional
systems is a major technical challenge, which has been overcome in a few cases
only. Most notably, for the cases studied here, a number of isolated periodic orbits
have been found in \cite{Kawahara2001,Viswanath2007,Cvitanovic2010}.
In the present case their identification is considerably
simplified by the presence of a period doubling cascade, that arranges them in families.

The outline of the paper is as follows: in section~\ref{sec:phasespace} we discuss the organization of the
state space near the lowest saddle-node bifurcation 
in plane Couette flow in our domain. In section~\ref{sec:symdyn} we discuss the characterization of the internal dynamics
and bifurcations in terms of a discrete map. We follow this in section~\ref{sec:pos} with a discussion
of the periodic orbits and their properties.

\section{\label{sec:phasespace}Bifurcations}
We study an incompressible Newtonian fluid between two infinitely extended plates in the $x$-$z$ plane 
located at $y=\pm h$ that move with speed $\pm U_0$ along the $x$-axis.
  A length-, velocity and time-scale are given by $h$, $U_0$ and $t=h/U_0$.
  The control parameter is the Reynolds number $\re=U_0d/\nu$, where $\nu$ is the viscosity of the fluid.
  We use periodic boundary conditions in the spatially extended directions and no-slip boundary conditions at the walls,
  with a box size of $2\pi\times2\times\pi$ in the downstream, wall-normal and spanwise directions, respectively.
  We use channelflow \cite{website:channelflow} to integrate this system with a Fourier-Chebyshev-Fourier scheme and a
  numerical resolution of $32\times33\times32$ modes; some key results, like the stability properties of our solutions, have been verified at a doubled resolution of $64\times65\times64$ modes.
  The main quantity we use in this study is the cross-flow energy, i.e.\ the amount of energy in the 
  flow components transverse to the laminar base flow,
  \begin{equation}
  \ecf = \frac 1V \int_V\left(v^2 + w^2\right)\mathrm dV,
  \end{equation}
  or the square root of $\ecf$ so as to reduce the range of it. 
  
  The box size in this study is just slightly longer and narrower than the cell studied in Ref. \cite{Waleffe1995}. 
  In this and other smaller boxes the typical flow consists of a pair of counter-rotating streamwise vortices and a 
  pair of alternating streaks.
  This limitation in the number of possible flow structures greatly simplifies the sequence of bifurcations.
  To simplify the analysis even further we enforce a shift-and-reflect symmetry
  \begin{equation}
    s_1[u,v,w](x,y,z) = [u,v,-w](x+L_x/2,y,-z),
  \end{equation}
  which, up to a discrete displacement by $L_z/2$, fixes the spanwise location of the flow structures.
  While we have not imposed further symmetries, all our solutions satisfy the additional symmetries
  \begin{equation}
    s_2[uvw](x,y,z) = [-u,-v,w](-x+L_x/2,-y,z+L_z/2)
  \end{equation}
  and
  \begin{equation}
    s_3[uvw](x,y,z) = [-u,-v,-w](-x,-y,-z+L_z/2),
  \end{equation}.
  These are the symmetries of the Nagata-Busse-Clever equilibrium, see e.g.\ \cite{Schmiegel1997,Gibson2008} for a discussion.
    
  The different box size compared to the calculations by Nagata \cite{Nagata1990} and 
  Clever and Busse \cite{Clever1997} has the effect
  of moving the first bifurcation to a higher Reynolds number, namely $\re=163.8$, 
  where a pair of fixed points appears in a saddle-node bifurcation. 
  The lower branch equilibrium (LB) has one unstable direction, 
  the upper branch equilibrium (UB) is linearly stable in the symmetry-reduced subspace.
  Since the laminar profile is linearly stable for all Reynolds numbers \cite{DrazinReid}, there now coexist two different kinds of attractors, the laminar flow and a set of attractors related to UB by symmetry.
  The separatrix of their basins of attraction is formed by the stable manifold of LB. 
  
  \begin{figure}
    \includegraphics{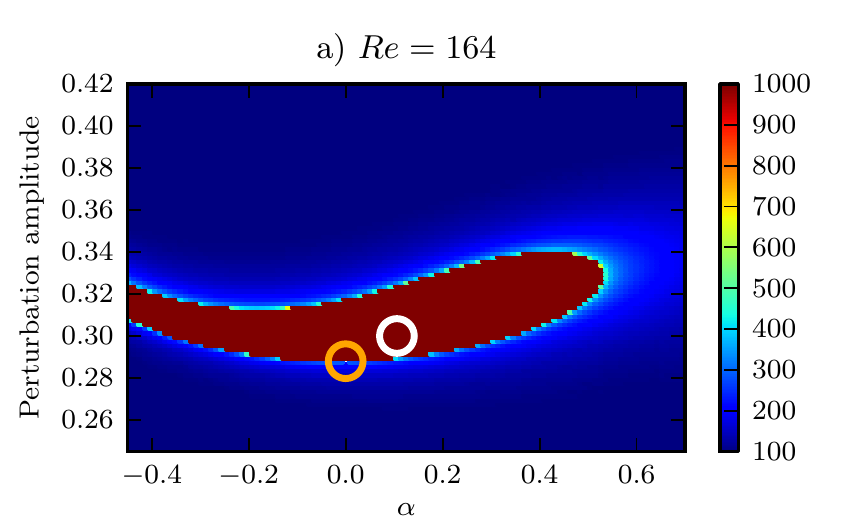}
    \includegraphics{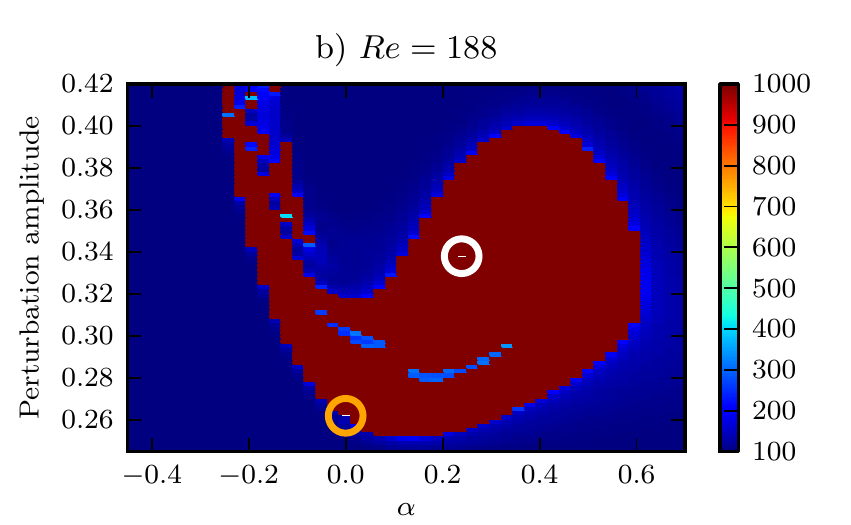}
    \includegraphics{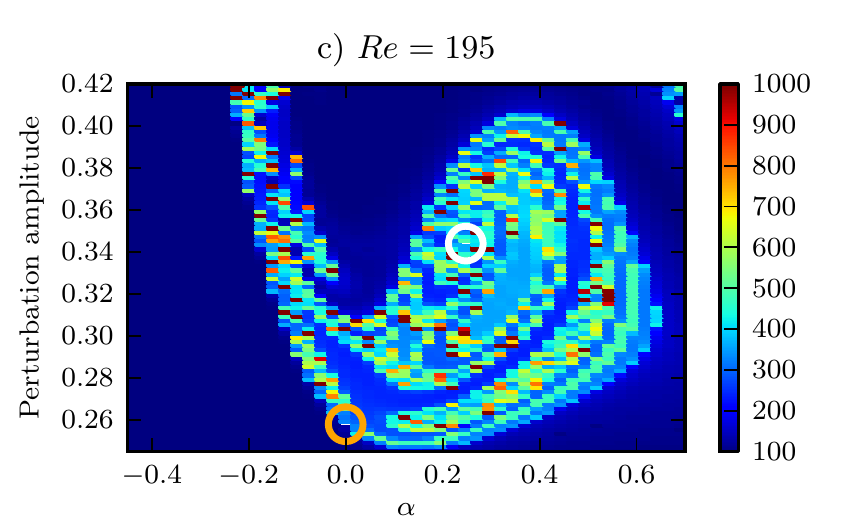}
    \caption{\label{fig:lifetimes}
      The basins of attraction of the attractors in plane Couette flow for three Reynolds numbers near the bifurcation point of the 
      Nagata-Clever-Busse state.
      Basins are depicted by the lifetime of initial conditions in advective timeunits, i.e.\ the time it takes for an initial condition to come into a distance $\ecf(u(t) - u_{lam}) < \epsilon$ of the laminar state.
      States on the $x$-axis are obtained by interpolating between the lower- and the upper-branch solution, rescaled by a factor that ensures that the distance between
      the two solutions in the plot is equal to the $L_2$-norm of their difference, see equation (\ref{eq:ualpha}).
      These states are then scaled in amplitude along the y-axis. The lower branch equilibrium is marked with an orange circle, the upper one with a white circle.
      (a) For $\re$ just slightly above the bifurcation point, the basin of attraction of the upper branch is a small region with a bubble-like shape. Its boundary is formed by the stable manifold of the lower branch.
      (b) For $\re=188$, the overall structure has not changed a lot, though the attractor is no longer a fixed point but a chaotic orbit.
      (c) At $\re=188.51$ a boundary crisis occurs -- the chaotic attractor becomes ``leaky'' and all trajectories eventually relaminarize. This bubble-like shape remains visible as an envelope of the initial conditions for longer living states, though.
    }
  \end{figure}
  The situation is illustrated in figure~\ref{fig:lifetimes}(a). The initial conditions that correspond to the points in the 
  plane of the figures are spanned by the laminar profile,
  and the upper and lower branch states. 
      The $x$-axis is a linear interpolation between LB and UB and the $y$-axis gives the perturbation amplitude with which
 the state is scaled. Formally, 
 the initial conditions are given by 
  \begin{equation}
  u(\alpha,A) = A\cdot u_{\alpha} / \|u_{\alpha}\|
  \end{equation}
  with 
  \begin{equation}\label{eq:ualpha}
  u_{\alpha} = u_{LB} + \frac{\alpha\cdot(u_{UB} - u_{LB})}{\sqrt{\|u_{UB} - u_{LB}\|^2 - (\|u_{UB}\|-\|u_{LB}\|)^2}}.
  \end{equation}
  The division by the root scales the $x$-axis in a way that the distance between LB and UB in the plot corresponds to the $L_2$-norm of the difference between the two fixed points.

  The basins of attraction are obtained by studying the lifetimes, defined as the time it takes for the flow
  to satisfy for the first time the condition $\ecf(u(t) - u_{lam}) < \epsilon = 10^{-8}$. Initially, when the basin of attraction
  is closed and trajectories are attracted to a stable fixed point that is not the laminar profile, there is a solid block of long,
  actually infinite, lifetimes. The stable fixed point lies somewhere in the center of the blob and the unstable saddle lies
  on its boundary. 
  
  A notable feature of these plots is the bubble shape of the basins of attraction of the stable fixed points.
  The unstable manifold of the unstable fixed point that loops around the stable fixed point
  folds back to approach the other branch of the stable manifold. This behaviour was identified by 
  Lebovitz \cite{Lebovitz2009,Lebovitz2012} in a model flow, and as the present figures shows it also is relevant to 
  full representations of the flow. In view of the bifurcation structures described in \cite{Mellibovsky2011} 
  one can expect that this bubble shape is rather generic.

  As Reynolds number increases, the stability properties of LB do not change up to at least $\re=O(10^4$ \cite{Wang2007}.
  On the other side, UB is only an attractor for a very short range and undergoes a Hopf bifurcation at $\re_H=166.05$, which has already been noted by Clever and Busse \cite{Clever1997}.
  The bifurcation is supercritical, as is revealed by the emergence of a limit cycle whose amplitude grows as $\sqrt{\re-\re_H}$.
  The emerging stable periodic orbit undergoes a period doubling at $\re=178.9$ which, for this domain size, results in a period doubling cascade.
  
\begin{figure}
    \includegraphics{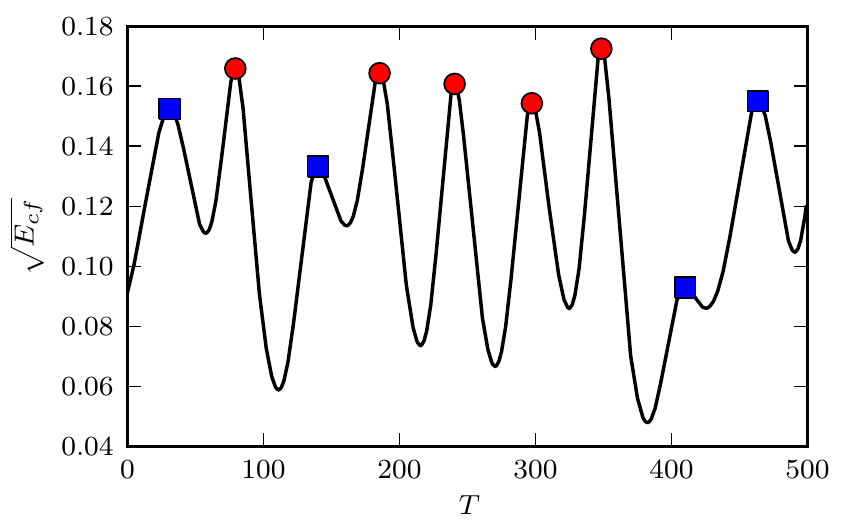}
    \caption{\label{fig:trajexample} Example of a typical chaotic trajectory at $\re=187.8$. For the reduction to a map, we only
    keep the energies at the maxima. The different symbols assigned to the maxima follow from the symbolic dynamics that will be introduced in section~\ref{sec:symdyn}: squares represent the symbol 0 and circles the symbol 1.}
  \end{figure}

  While the way to represent fixed points in a bifurcation diagram is straightforward, it is less so for periodic orbits or chaotic trajectories.
  To visualize the attracting state once it gets more complicated we introduce a reduced representation of the
  flow by focussing on the maxima in energy: we calculate a typical trajectory originating near UB
  (and hence not lying in the immediate basin of attraction of the laminar state), omit an initial transient and all intermediate
  times except for the maxima. We then keep values of the maxima in  $\sqrt{\ecf}$ to represent the flow.
The method is illustrated in 
figure~\ref{fig:trajexample}, where a short section of a typical chaotic trajectory for $\re=187.8$ is shown. 
The maxima are marked with symbols corresponding to the symbolic dynamics that will be introduced in section~\ref{sec:symdyn}.
  
  \begin{figure}
    \includegraphics{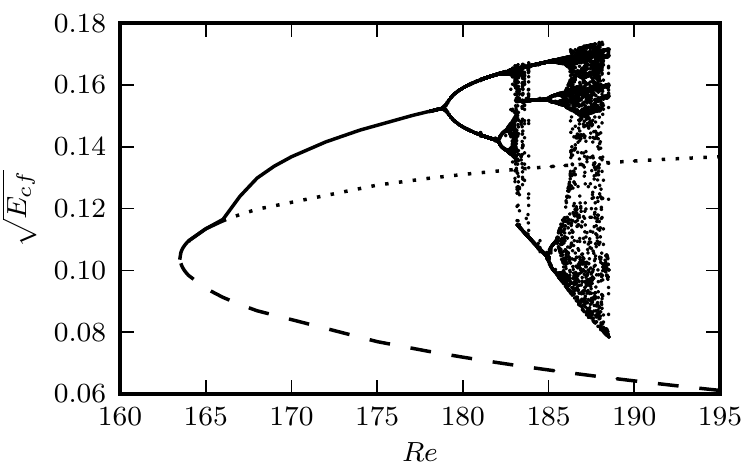}
    \caption{\label{fig:bifdiag}
      The square root of the cross-flow energy $\sqrt{\ecf}$ for the upper- and lower-branch solutions and the attractor coexisting with the laminar state.
      To visualize periodic orbits and chaotic attractors, we plot every maximum of $\ecf$ with a dot.
      The Nagata-Busse-Clever state is born in a saddle-node bifurcation at $\re=163.8$, we plot the lower branch with a dashed  line and the upper branch with a line as long as it is stable, and with a dotted line afterwards.
      The upper branch loses stability in a Hopf bifurcation at $\re=166.05$, the emergence of a limit cycle is nicely visible. It undergoes a period doubling cascade.
      After the boundary crisis at $\re=188.51$, the laminar state is the only attractor.
    }
  \end{figure}

  The bifurcation cascade is summarized in figure~\ref{fig:bifdiag}, where the square root of the cross-flow energy  $\ecf$ is plotted.
  The attractor is plotted with a dot for every maximum of $\ecf$, LB is a dashed line and UB is a dotted line once it becomes unstable.
  In this representation the emergence of a limit cycle at $\re_H$ with an amplitude that grows as $\sqrt{\re-\re_H}$ is clearly visible.
  It is followed by a period doubling cascade resulting in a chaotic attractor.
  We find a stable period three window from $\re=183.8$ to $\re=185.0$, followed by a new period doubling cascade of the period three orbit.
    The basin of attraction is still confined to a small region in state space, as is evident from figure~\ref{fig:lifetimes}(b), which shows the basin for $\re=188$.
  Compared to the above picture at $\re=164$, the general shape of the basin has not changed. Near the crisis bifurcation perhaps
  also the homoclinic tangles described in \cite{vanVeen2011} form. The transition in the lifetimes is in agreement with 
  the phenomenology observed also in the two-dimensional map described in \cite{Vollmer2009}.
    
  For $\re > \re_c = 188.51$, there are no more dots in figure~\ref{fig:bifdiag}: all trajectories eventually relaminarize.
  Our interpretation is that at $\re_c$ the chaotic attractor touches the fixed point LB (or, equivalently, its stable manifold) and a boundary crisis \cite{Grebogi1983} occurs and the attractor becomes ``leaky''. 
  Even though the crisis is not visible in figure~\ref{fig:bifdiag} due to the chosen representation, it will become visible later in another view of state space, see figures \ref{fig:id} below.
  We show a state space plot of $\re=195$ in figure~\ref{fig:lifetimes}(c).
  The bubble shape that formed the basin of attraction of the chaotic attractor before the crisis is still visible.
  But the lifetimes of the initial conditions from that region now are no longer infinite but vary drastically between almost direct decay and very long survival.
  The distribution of the lifetimes (not shown here) turns out to be  exponential -- a signature of a chaotic saddle left behind by the boundary crisis \cite{Lai2011}.

\section{\label{sec:symdyn}Symbolic dynamics}

\begin{figure}
    \includegraphics{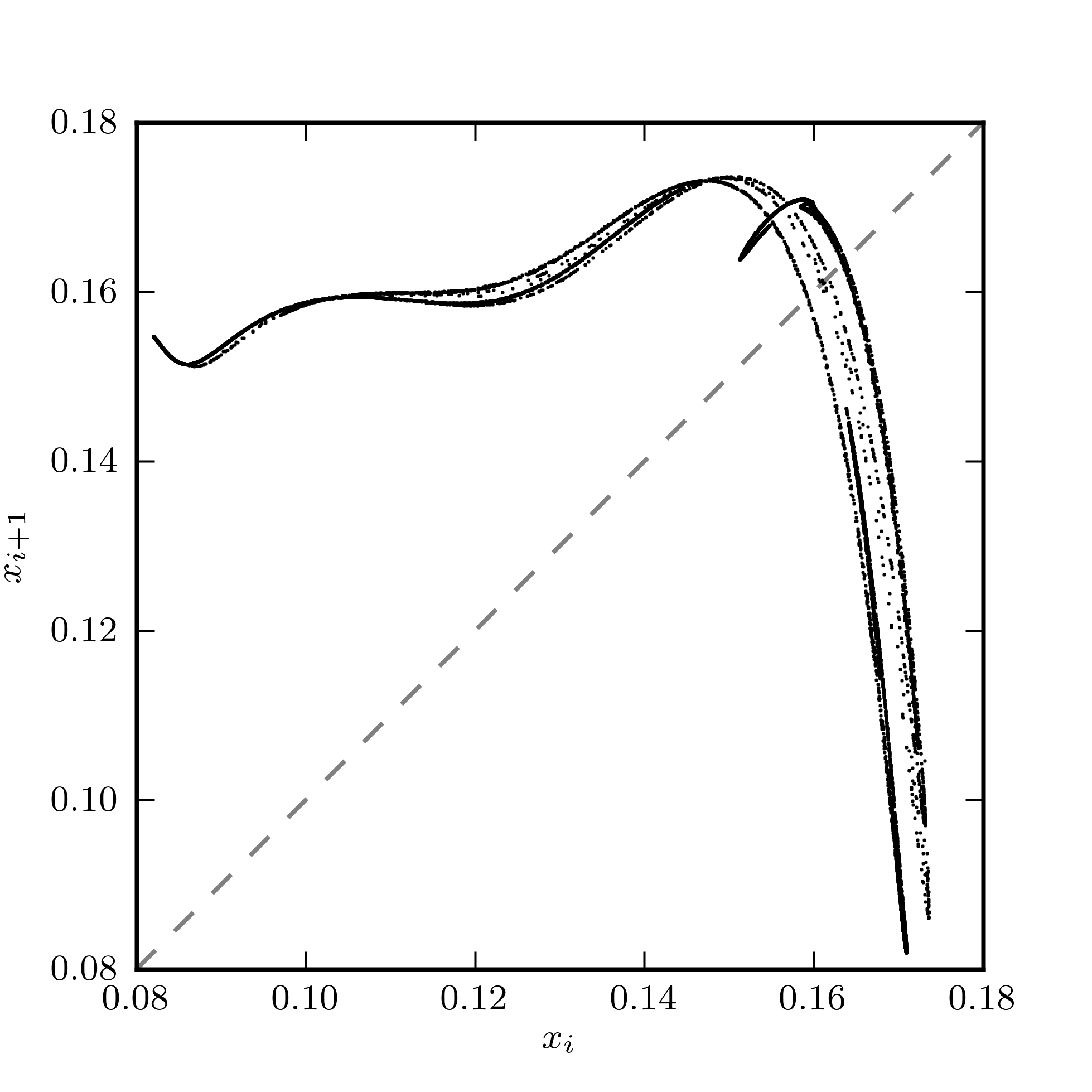}
    
    \caption{\label{fig:attractor}
      The chaotic attractor at $\re=187.8$. It is obtained by calculating for a trajectory that contains 10799 maxima of $\ecf$.
      The values of the maxima of $\sqrt{\ecf}$ are denoted $x_i$ for the i-th maximum and the map is obtained by
      plotting $x_{i+1}$ vs. $x_i$.
      It consists of several branches that lie close to each other and, for the purposes of labelling trajectories, can 
      be collapsed into a one-dimensional uni-modal map.
      A binary symbolic dynamics can be introduced according to the relative location of the points:
      points that lie on a branch left or right of the maximum are labeled $0$ or $1$, respectively.
    }
  \end{figure}

  In this and the next section we focus on the situation just before the crisis, namely at $\re=187.8$,
where the attractor appears chaotic. We now use the reduction of the flow to the location of its maxima to introduce
an approximately one-dimensional map of the flow.
We denote the value of the $i$-th maximum of $\sqrt{\ecf}$ along this trajectory by $x_i$ and plot
$x_{i+1}$ vs. $x_i$. This then gives a 1-dimensional projection of the Poincar\'e return map, $x_i \rightarrow x_{i+1} = f(x_i)$.
  Figure~\ref{fig:attractor} shows a plot of the map obtained from a trajectory with $10798$ maxima.

  In this representation, the attractor looks very thin and almost one-dimensional, though a fractal structure can be seen perpendicular to it.
  In this sense it is qualitatively similar to fractal attractors in well-known low-dimensional systems like the H\'enon map with standard parameters.
  The attractor consists of several branches, all of which show a distinctive maximum between $x_i=0.14$ and $x_i=0.16$, but
  the location of the maxima do not coincide.

  \begin{figure}
    \includegraphics{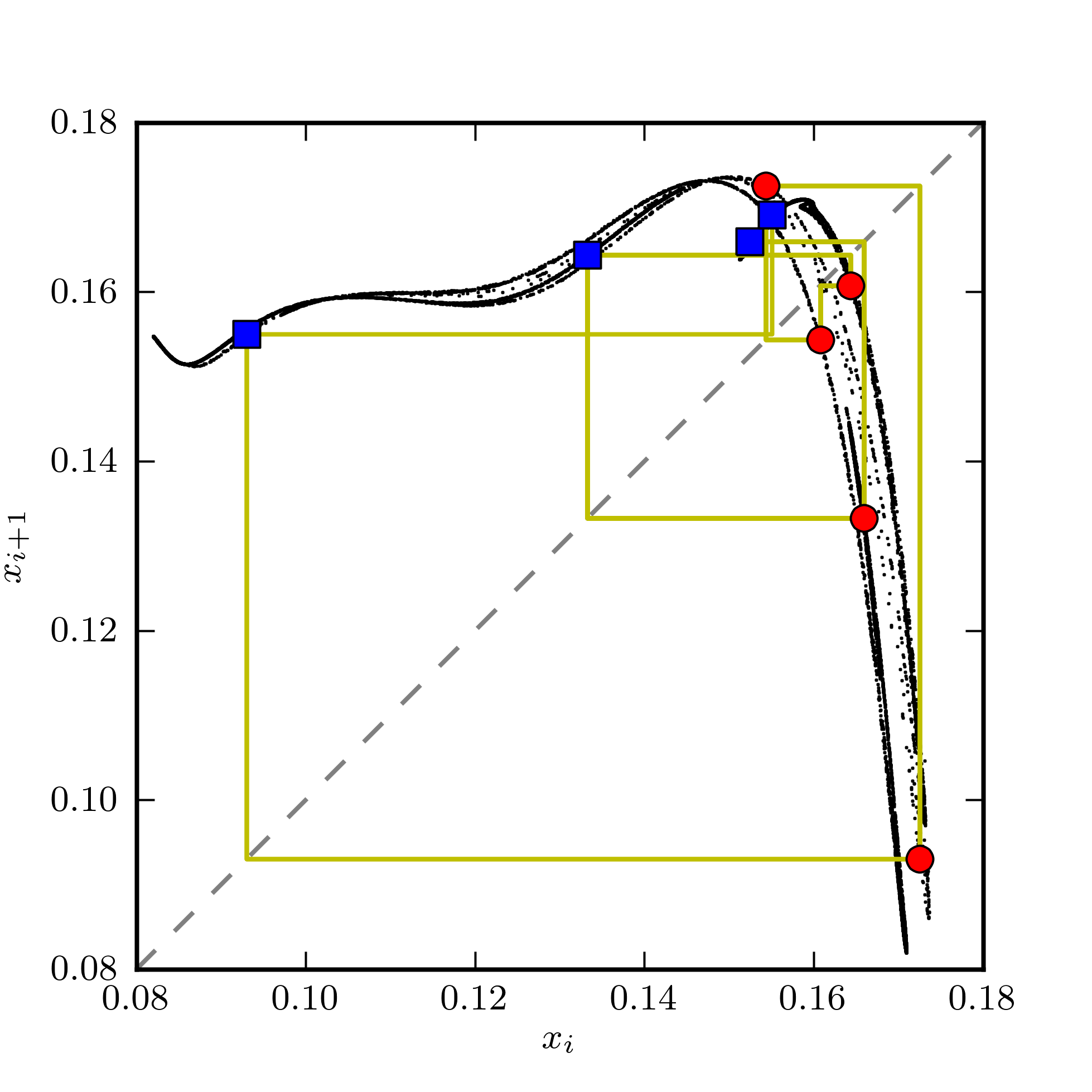}
    \caption{\label{fig:attexample}
      Symbolic dynamics for the same trajectory as in figure~\ref{fig:trajexample}, squares are $0$ and circles $1$.
      The rightmost square is left of the maximum and has to be labeled as $0$ despite the fact that it is further right than the leftmost circle.
    }
  \end{figure}
  Given this representation of the attractor, we introduce a binary symbolic dynamics in the usual way: on each branch, a point is labeled as $0$ or $1$ if it lies to the left or right of the maximum, respectively.
  The symbolic dynamics is shown in figure~\ref{fig:attexample} for the same snapshot of the trajectory as in figure~\ref{fig:trajexample}.
We find that a partition cannot be based on a single value of $x_i$ and that the branch on which it comes to
lie also has to be taken into account: $x_6$ is a little lower than $x_9$ but assigned the symbol 1
while $x_9$ is assigned 0 because of its different location relative to the maximum of the corresponding branch. 
The two rightmost squares in figure~\ref{fig:attexample}(b) lie on the branch of the rightmost maximum while the circle 
just above them lies on the branch of a maximum further left.

\section{\label{sec:pos}Periodic orbits}

  \begin{table*}
    \begin{ruledtabular}
      \begin{tabular}{rrrrrcrrr}
        Orbit  & $T$ &  $\re_{Bif}$ & $\Lambda_1$ & $\lambda_1$ & $\Lambda_2 - 1$    & $\Lambda_3 - 1$      & $\Lambda_4$ & $\lambda_4$ \\
        $\overline{1}$ & $55.22$ & $166.0$ & $-2.64$ & $0.0176$ & $9.83\cdot 10^{-7}$ & $-5.65\cdot 10^{-7}$ & $0.4$ & $-0.0166$ \\
        $\overline{01}$ & $106.95$ & $179.3$ & $-4.79$ & $0.0146$ & $2.11\cdot 10^{-6}$ & $-5.33\cdot 10^{-7}$ & $0.13$ & $-0.0191$ \\
        $\overline{001}$ & $162.1$ & $181.9$ & $-4.6$ & $0.0094$ & $1.95\cdot 10^{-5}$ & $2.76\cdot 10^{-6}$ & $0.05$ & $-0.0185$ \\
        $\overline{011}$ & $159.21$ & $181.9$ & $9.43$ & $0.0141$ & $1.37\cdot 10^{-7}$ & $-4.33\cdot 10^{-6}$ & $0.048$ & $-0.0191$ \\
        $\overline{0111}$ & $215.84$ & $182.1$ & $-27.73$ & $0.0154$ & $3.56\cdot 10^{-4}$ & $-8.81\cdot 10^{-7}$ & $0.017$ & $-0.0189$ \\
        $\overline{01101}$ & $266.87$ & $182.2$ & $-50.67$ & $0.0147$ & $4.21\cdot 10^{-3}$ & $-1.70\cdot 10^{-5}$ & $0.0063$ & $-0.0190$ \\
        $\overline{01111}$ & $270.65$ & $182.2$ & $76.63$ & $0.0160$ & $4.87\cdot 10^{-6}$ & $4.87\cdot 10^{-6}$ & $0.0061$ & $-0.0188$ \\
      \end{tabular}
    \end{ruledtabular}
    \caption{\label{tab:properties}
      Properties of the periodic orbits.
      The first column contains the symbolic sequence of the orbit, the second column the period in advective time-units of Navier-Stokes. 
      The third column gives the Reynolds numbers where the orbits are created.
      The remaining columns give the eigenvalues with the largest magnitudes.
      where $\Lambda_i$ denotes the eigenvalues of the monodromy matrix and $\lambda_i=\ln |\Lambda_i| / T$ the Lyapunov exponents.
      The eigenvalues $\Lambda_2$ and $\Lambda_3$ correspond to streamwise and spanwise translations and are mathematically equal to one, the difference $\Lambda-1$ serves as a measure for the accuracy of the numerics.
      All orbits have one and only one unstable eigenvalue.
    }
  \end{table*}

  In this section, we will describe the periodic orbits we found at $\re=187.8$.
    We obtain initial guesses by scanning the map $f$ (figure~\ref{fig:attractor}) for close returns along a trajectory and then use channelflows excellent Newton-Krylov-Hookstep algorithm, based on \cite{Viswanath2007} and implemented by John F Gibson in channelflow \cite{website:channelflow}, for finding periodic orbits.
  Since the attractor for this Reynolds number is chaotic, we expect all orbits to be linearly unstable.
  For reasons of computational stability we restrict the search to orbits of symbolic period not exceeding five.
  
\begin{figure}
    \includegraphics{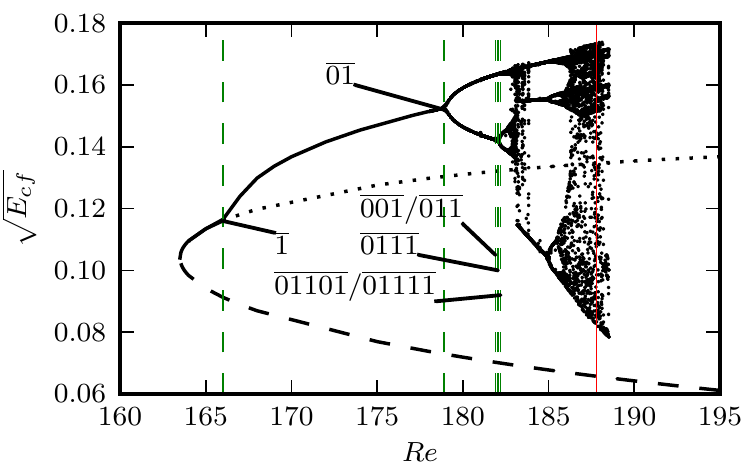}
    \caption{\label{fig:bifdiag2}
      Critical Reynolds numbers for the periodic orbit bifurcations, indicated with dashed lines. 
      The straight line at $\re=187.8$ corresponds to the Reynolds number where the analysis takes place.
    }
  \end{figure}

  With this method we found seven different orbits -- they are listed, along with their properties in table~\ref{tab:properties}.
  The first column contains the symbolic name of the orbit, the second one the period of the orbit in time-units of Navier-Stokes.
    The Reynolds number where the orbits bifurcate, either pitchfork or saddle node bifurcations, are given in the fourth column. 
  The orbit $\po{1}$ is created in a Hopf-bifurcation of UB, the orbits $\po{01}$ and $\po{0111}$ result from period-doublings of the former one.
  The pairs of the two period-3 and period-5 orbits are the result of saddle-node bifurcations.
  To better illustrate the bifurcation sequence, we show the bifurcation diagram again in figure~\ref{fig:bifdiag2} with the bifurcation points indicated by dashed lines. The solid line at $\re=187.8$ indicates the Reynolds number where the analysis of the symbolic dynamics takes place.
  
  The bifurcation sequence of the orbits is almost the same as the universal sequence described by Metropolis, Stein and Stein \cite{Metropolis1973}, with the only exception being that the period-doubling of $\po{01}$ that creates $\po{0111}$ takes place slightly after the saddle-node bifurcations in which the orbits $\po{001}$ and $\po{011}$ appear.
  We have not found the orbits $\overline{0011}$, $\overline{0001}$, $\overline{00111}$, $\overline{00011}$, $\overline{00001}$ and 
  $\overline{00101}$ -- which is in accordance with the universal sequence and also found in the the logistic map for $R<4$.

  The eigenvalues $\Lambda_i$ of the monodromy matrix 
  of the periodic orbits are calculated using channelflow's Arnoldi iteration \cite{Viswanath2007}.
  From the eigenvalues $\Lambda_i$ one can extract Lyapunov exponents $\lambda_i=\ln |\Lambda_i| / T$. 
  The four eigenvalues of largest magnitude and the corresponding nontrivial Lyapunov exponents are given in the last 
  six columns of table~\ref{tab:properties}.
  All of the periodic orbits have exactly one unstable eigenvalue $\Lambda_1$ and two eigenvalues $\Lambda_{2,3}$ that correspond to translation along $x$ and time that are theoretically strictly equal to $1$ (translations along $z$ are excluded by the enforced shift-and-reflect symmetry). The deviation of $\Lambda_{2,3}$ from $1$ can be used as a measure of the accuracy of our numerics. If the calculations are restricted to the full symmetry-subspace of the Nagata state, the marginal eigenvalues corresponding to shifts in the downstream direction disappear.
  $\Lambda_4$ is the least stable eigenvalue of the orbit and measures the contraction in the directions perpendicular to the attractor.
  The positive first Lyapunov exponent varies between $0.0094$ and $0.176$, where the largest value comes from the shortest orbit $\po1$.
  All orbits have exactly one unstable direction and their weakest contracting Lyapunov exponent $\lambda_4$ is of the same order,
  this supports the conjecture that the system can be approximated by a one dimensional map.
  
  The upper branch becomes unstable via a Hopf bifurcation and at $\re=187.8$ the complex conjugate pair of eigenvalues is still the only unstable ones.
  The values are $\lambda_{1,2} = 0.023\pm i0.128$ and the imaginary part corresponds to a period of $T = 2\pi/\mbox{Im}(\lambda) = 48.93$ which is not too far from the period of the orbit $\po{1}$.
  
  We have not been able to determine the orbit $\po{0}$.

\begin{figure}
    \includegraphics{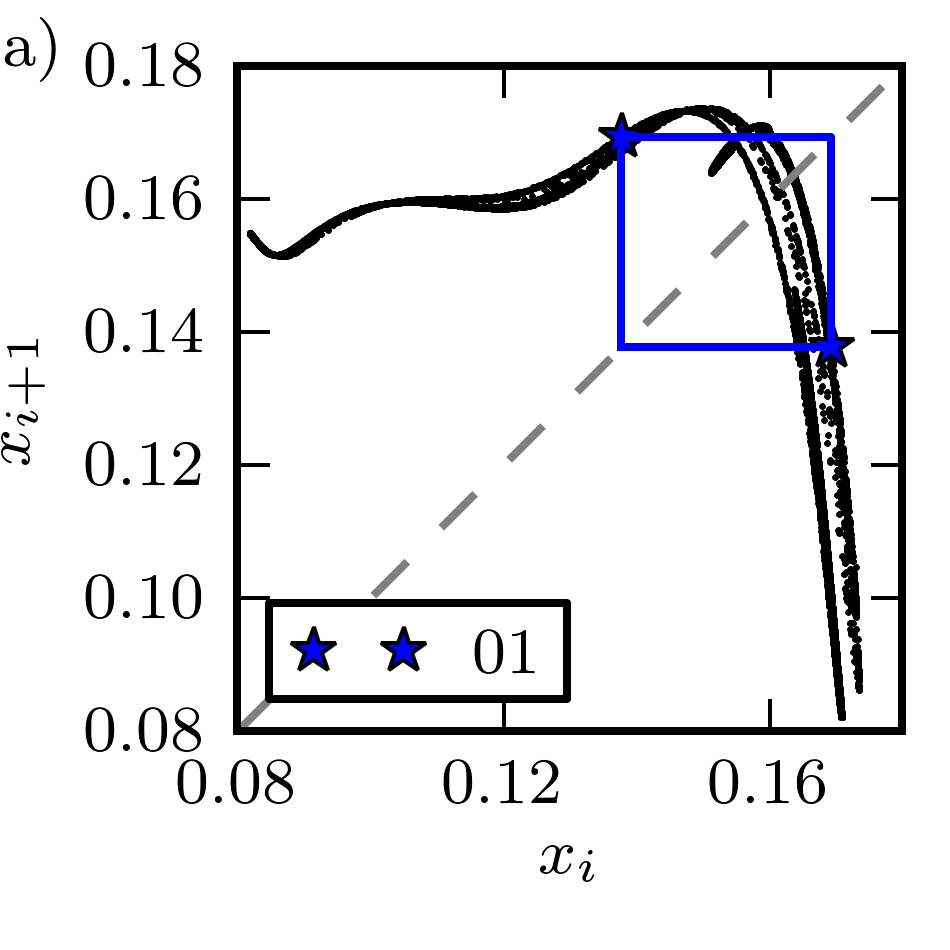}
    \includegraphics{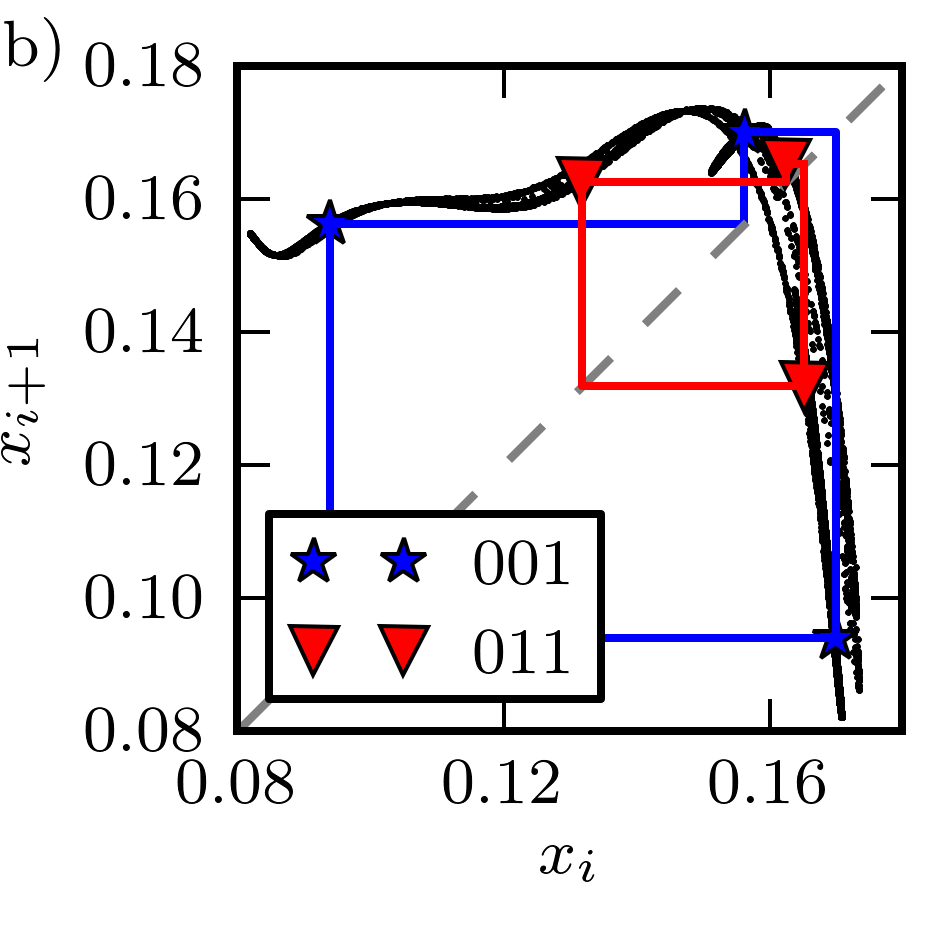}
    
    \includegraphics{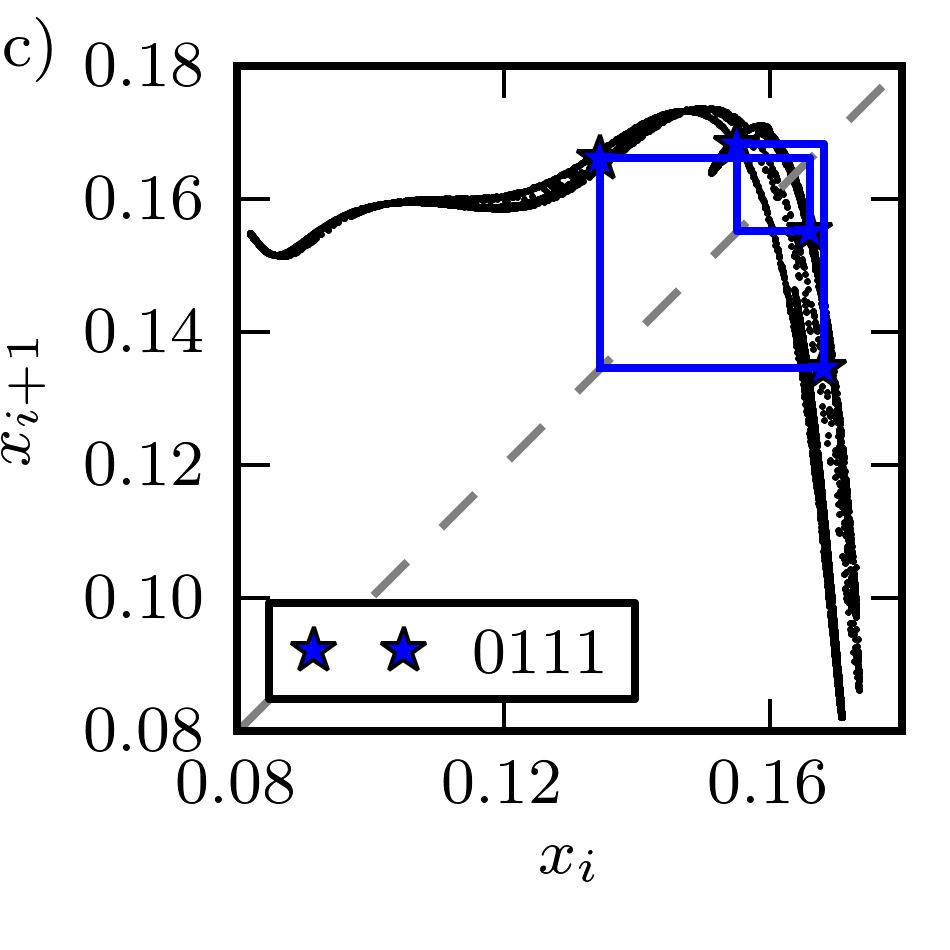}
    \includegraphics{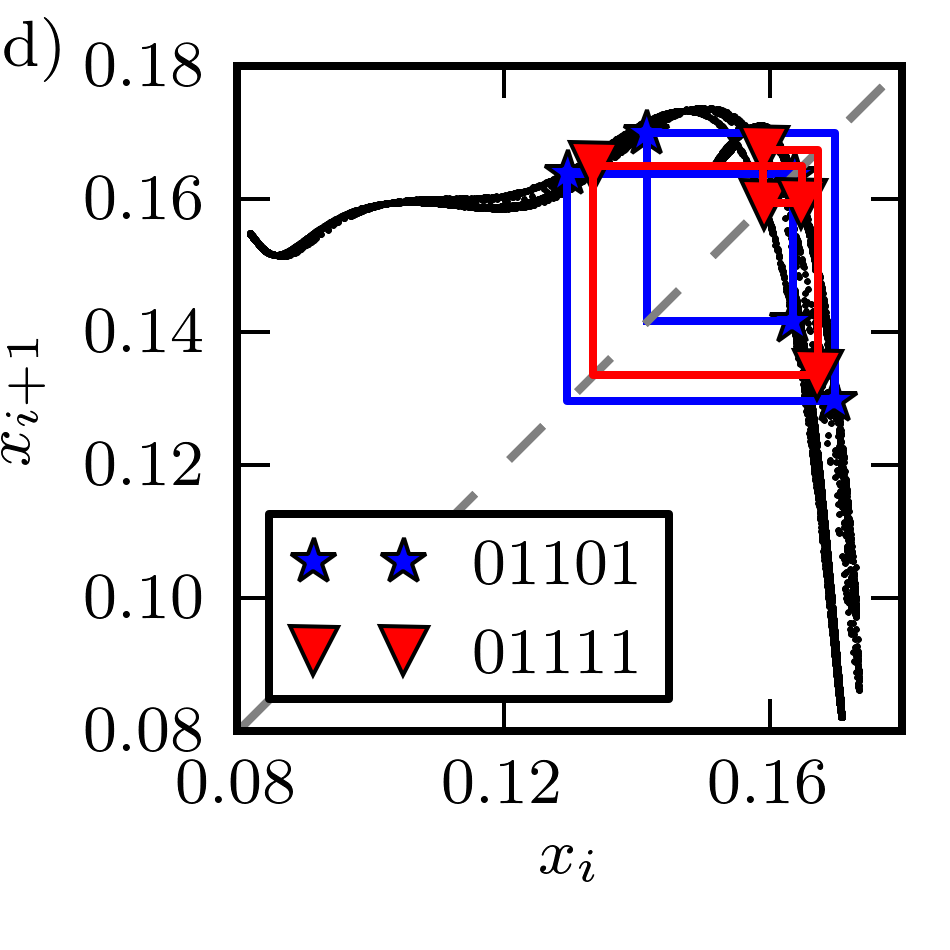}
    
    \caption{\label{fig:orbits}
      The periodic orbits, presented in cobweb-plots. Only the orbit $\po{001}$ visits the extremer position on the attractor.
    }
  \end{figure}

  Cobweb-plots of all orbits are presented in figure~\ref{fig:orbits}. While most of the orbits concentrate in the upper right corner of the map, only $\po{001}$ stretches out to the lowest and highest values. Since $\po{001}$ is the only orbit with a sequence $00$, this suggests that longer orbits containing $00$ would also stretch out to those more extreme regions of the attractor.

\begin{figure}
    \includegraphics[width=\linewidth]{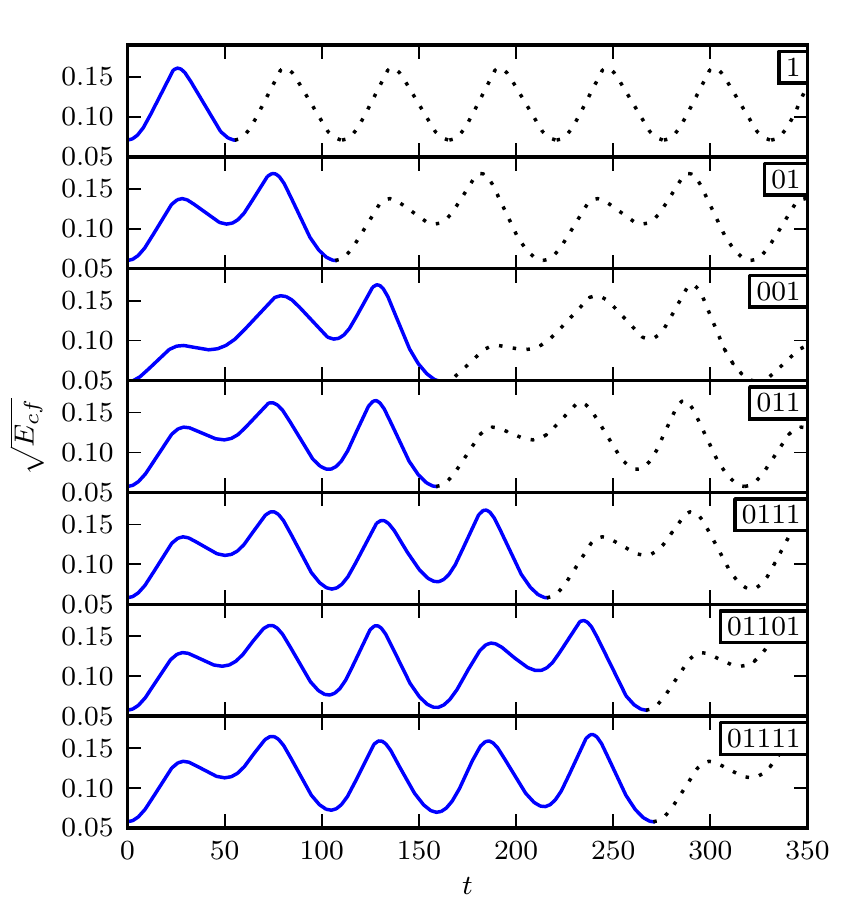}
    \caption{\label{fig:ecf}
      The root of the cross-flow energy $\sqrt{\ecf}$ for all periodic orbits. One cycle of each periodic orbit is shown solid, with continuation dotted. The composition of the longer orbits from parts of the shorter ones is nicely visible.
    }
  \end{figure}

\begin{figure}
    \includegraphics[width=\linewidth]{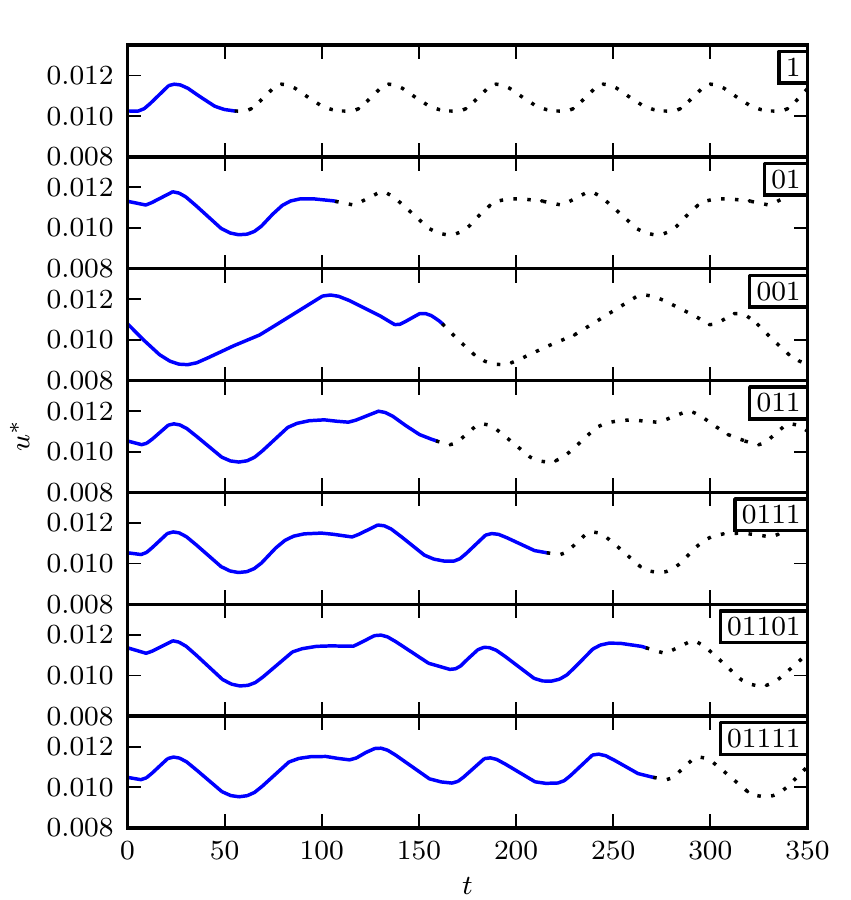}
    \caption{\label{fig:ustar}
      The wall shear rate for all periodic orbits. One cycle of each periodic orbit is shown solid, with continuation dotted. While $u^*$ has maxima at roughly the same points as $\ecf$, the symbolic dynamics can not be recognized.
    }
  \end{figure}

  In figures~\ref{fig:ecf} and~\ref{fig:ustar} we present orbits in a plane spanned by the cross-flow energy $\ecf$ and the wall shear rate $u^* = \sqrt{{\partial}/{\partial y} U|_{y=\pm1}}$, respectively.
  Figure~\ref{fig:ecf} shows nicely the composition of the longer orbits as repetitions of the shorter ones. For example, the orbit $\overline{011}$ is very similar to $\overline{0111}$ and $\overline{01111}$, except that the latter ones have one (two) additional peak(s) that looks quite similar to $\overline{1}$. Note also how the orbits explore the ranges covered by a turbulent trajectory,
  lending support to the idea that periodic orbits can be used to represent all possible motions and that averages 
  could be calculated from the periodic orbits.

\begin{figure}
    \includegraphics[width=\linewidth]{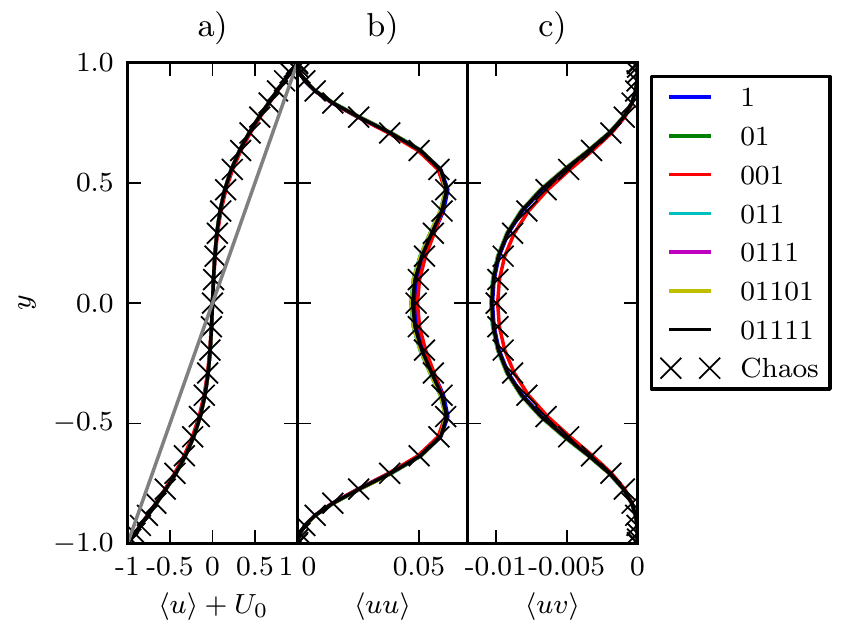}
    \caption{Averaged quantities of the periodic orbits (lines) and a chaotic trajectory (crosses) as a function of y. (a) Averaged downstream velocity u.(b) Fluctuations in the downstream component $\langle uu\rangle$. (c) Average of $\langle uv \rangle$.}
    \label{fig:orbavgs}
  \end{figure}

    We show some averaged quantities of the orbits in figure~\ref{fig:orbavgs}. We use small letters for the fluctuating components 
  and $U_0$ for the base flow.
  The left plot shows $\langle u \rangle + U_0$, the middle one $\langle uu \rangle$ and the right one $\langle vv \rangle$.
  In all three plots, the mean profiles for nearly all orbits are impossible to distinguish among each other. 
  Within the group of orbits of period less than six, the exception is $\po{001}$, which is marginally offset to the right in (b) and (c).
  We indicate the averages obtained from a chaotic trajectory with 5000 time units by crosses -- they can hardly be distinguished from the averages of the periodic orbits.

  The motion of the walls injects energy into the fluid motion which in turn is dissipated by viscosity. The energy balance can be written as \cite{Gibson2008}
  \[\dot E = I - D,\]
  where the change of the kinetic energy density $E$ is from the bulk viscous dissipation rate $D$ and the wall-shear power input $I$,
  \[E = \frac 1V \int_V \frac 12 \left(\vec{u} + \vec U_0\right)^2\mathrm dV,\]
  \[D = \frac 1V \int_V \left( \vec \nabla \times (\vec{u} + {\vec U}_0)\right)^2\mathrm dV,\]
  \[I = 1 + \frac 1{2A}\int_A\left(\left.\frac{\partial u}{\partial y}\right|_{y=1} + \left.\frac{\partial u}{\partial y}\right|_{y=-1}\right)\mathrm dx\mathrm dz.\]
  Here $V = 2 L_x L_z$ and $A =  L_x L_z$. The normalizations are chosen so that
    $D = I = 1$ for laminar flow and $\dot{E} = I - D$.
    
  \begin{figure}
    \includegraphics{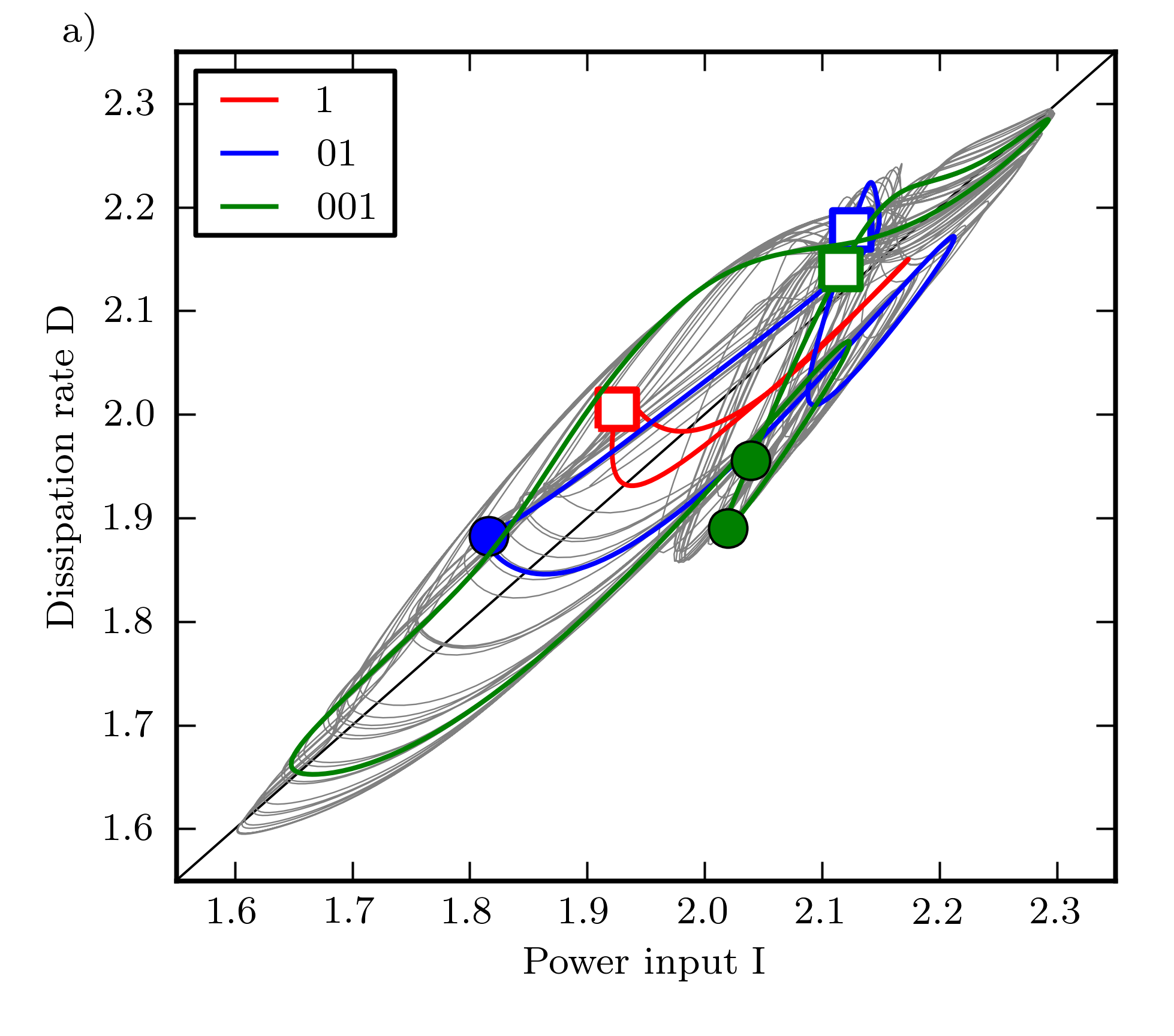}\\
    \includegraphics{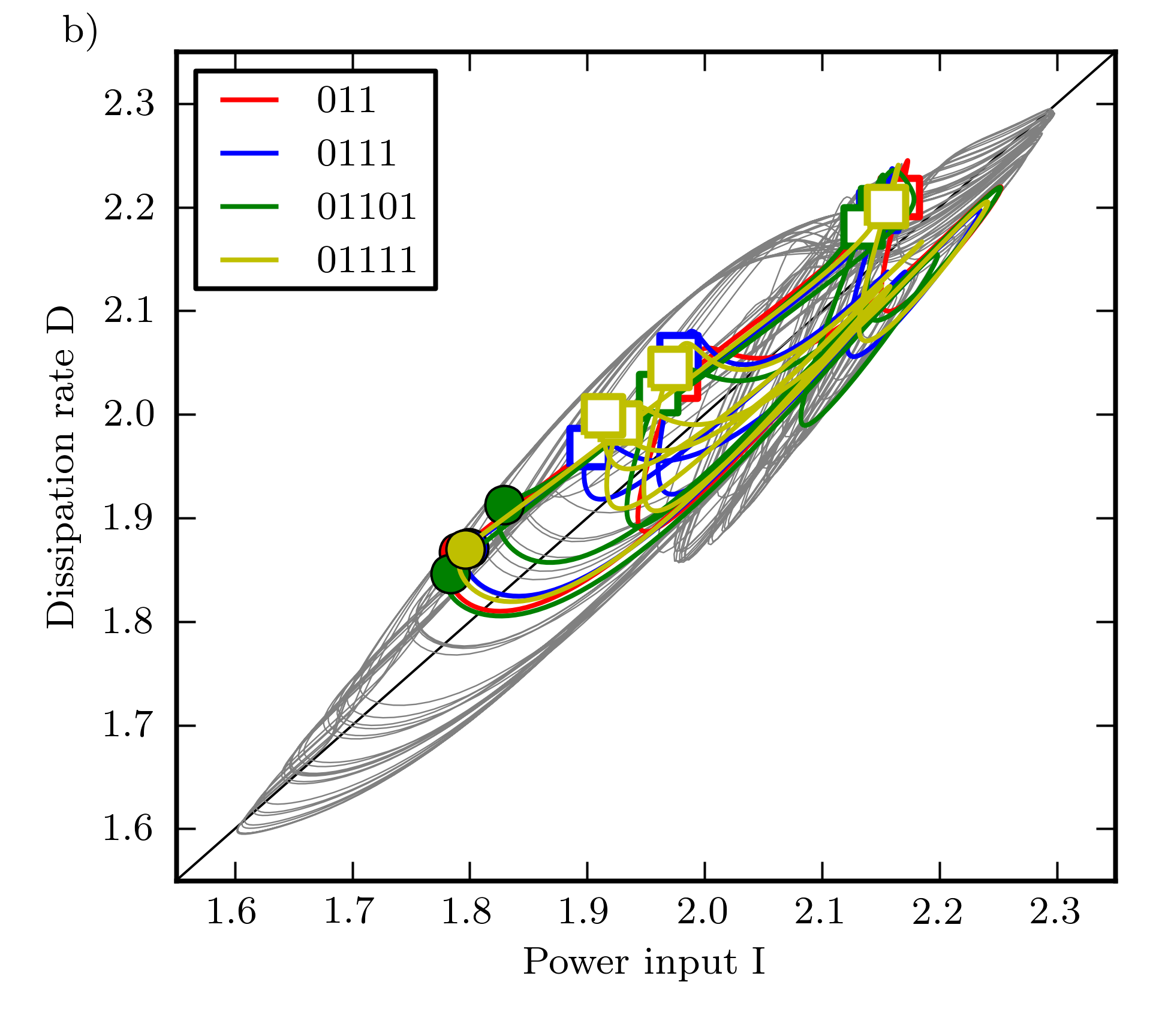}
    \caption{\label{fig:id}
      The periodic orbits in the input-dissipation plane. In the background a chaotic trajectory is plotted in gray. 
      The maxima of $\ecf$ along the orbits are indicated with circles for symbol $0$ and open squares for $1$. 
      In this plots, the regions of the different symbols do not overlap.
      It is interesting to notice that the only two symbols below the diagonal belong to the orbit $\po{001}$.
    }
  \end{figure}
  
  In a plot of $I$ vs. $D$, fixed points must lie on the diagonal $I=D$. For periodic orbits and also for the
  non-periodic chaotic motions, the center,
  defined as time-averages of $I$ and $D$, will also lie on the diagonal. 
  In figure~\ref{fig:id} we present all periodic orbits and a chaotic trajectory.
  The points corresponding to the maxima of $\ecf$ are marked with circles for symbol $0$ and squares for $1$,
  the diamond shaped symbol in the lower left corner marks LB.
  There are three points noteworthy in this figure:
  First, the symbols are clearly separated and a partition could be based on the location of a maximum in that representation.
  Second, the periodic orbits capture all of the qualitative features and loops of the chaotic trajectory.
  And third, one sees that the chaotic trajectory in the background is about to touch LB, the event that will trigger the crisis.

\section{Final remarks}
The detection of periodic orbits in this system was certainly assisted by the fact that the flows
studied here were confined to a small domain, restricted by a discrete shift and reflect symmetry,
and also analyzed at a low Reynolds number. In this sense, it is similar to the study of fluctuations 
in the Kuramoto-Shivashinsky equation \cite{Christiansen1997}. Nevertheless, the reduction to a symbolic dynamics compatible
with a unimodal one-dimensional map and the association of the orbits to the universal
Metropolis-Stein-Stein sequence shows that also this high-dimensional system follows the universal
dynamics identified in low-dimensional systems. This raises the hope that bifurcation analyses
and periodic orbits can also be identified and used in other systems, such as pipe flows or plane 
Poiseuille flow. The ultimate goal, however, is to analyze not only chaotic but also some turbulent
flows in terms of periodic orbits, as suggested e.g. in \cite{vanVeen2006}.

\section*{Acknowledgements}
We would like to thank John F Gibson for developing and providing access to his open source Channelflow.org code, on 
which the calculations presented here are based,
and Norman Lebovitz and Tobias M Schneider for helpful discussions.
We thank Predrag Cvitanovi\'c for maintaining a continously updated source on periodic orbit theory at chaosbook.org.
This work was supported in part by the German Research Foundation within Forschergruppe 1182.

\bibliographystyle{ieeetr}
\bibliography{library.bib}

\end{document}